\documentclass[english,12pt]{article}
\usepackage{array}
\usepackage{graphicx}
\usepackage{amssymb}
\usepackage[english]{babel}
\usepackage{amsmath}
\usepackage{multirow}
\usepackage{prettyref}
\usepackage{babel}
\usepackage{units}
\usepackage[latin1]{inputenc}
\usepackage{amsfonts}
\usepackage{amssymb}
\usepackage{babel}
\usepackage{color}

\usepackage{amsfonts}

\usepackage{authblk}
\usepackage{cite}
\def\@fmsl@sh#1#2#3{\m@th\ooalign{$\hfil#1\mkern#2/\hfil$\crcr$#1#3$}}
 \def\eq#1\en{\begin{equation}#1\end{equation}}
\def\s[#1,#2]{[#1\stackrel{\star}{,}#2]}
\def\sx[#1,#2]{[#1\stackrel{\star_{x}}{,}#2]}

\newcommand{\nc}{\newcommand}
\nc{\beq}{\begin{equation}}
\nc{\eeq}{\end{equation}}
\nc{\beqa}{\begin{eqnarray}}
\nc{\eeqa}{\end{eqnarray}}

\def\bc{\begin{center}}
\def\ec{\end{center}}

\def\to{\rightarrow}

\def\gsim{\mathrel{\mathpalette\atversim>}}

\def\bc{\begin{center}}
\def\ec{\end{center}}

\newcommand{\uunderline}[1]{\underline{\underline{#1}}}

\def\gsim{\mathrel{\rlap{\lower4pt\hbox{\hskip1pt$\sim$}}

    \raise1pt\hbox{$>$}}}       

\def\gsim{\mathrel{\rlap{\lower4pt\hbox{\hskip1pt$\sim$}}
    \raise1pt\hbox{$>$}}}       

\begin{document}
\makeatletter
\def\fmslash{\@ifnextchar[{\fmsl@sh}{\fmsl@sh[0mu]}}
\def\fmsl@sh[#1]#2{%
  \mathchoice
    {\@fmsl@sh\displaystyle{#1}{#2}}%
    {\@fmsl@sh\textstyle{#1}{#2}}%
    {\@fmsl@sh\scriptstyle{#1}{#2}}%
    {\@fmsl@sh\scriptscriptstyle{#1}{#2}}}
\def\@fmsl@sh#1#2#3{\m@th\ooalign{$\hfil#1\mkern#2/\hfil$\crcr$#1#3$}}
\makeatother

\thispagestyle{empty}
\begin{titlepage}
\boldmath
\begin{center}
  \Large {\bf   The Spectrum of Quantum Gravity}
    \end{center}
\unboldmath
\vspace{0.2cm}
\begin{center}
{  {\large Xavier Calmet}\footnote{x.calmet@sussex.ac.uk}$^{a}$}{\large and}  
{  {\large Boris Latosh}\footnote{b.latosh@sussex.ac.uk}$^{a,b}$} 
 \end{center}
\begin{center}
$^a${\sl Department of Physics and Astronomy, 
University of Sussex, Brighton, BN1 9QH, United Kingdom
}\\
$^b${\sl Dubna State University,
Universitetskaya str. 19, Dubna 141982, Russia}
\end{center}
\vspace{5cm}
\begin{abstract}
\noindent
In this paper we consider the degrees of freedom beyond the graviton present in the effective field  theory for quantum gravity. We point out that the position of the poles due to $R^2$ and $R_{\mu\nu}R^{\mu\nu}$ cannot be affected by operators that are higher order in curvature. On the other hand, operators of the type $R\Box R$ will lead to new poles while shifting the positions of the poles found at second order in curvature. New degrees of freedom can be identified either, as just described, by looking at the poles of the graviton propagator corrected by quantum gravity or by mapping the Jordan frame theory to the Einstein frame theory. While this procedure is very well defined for second order curvature terms in the effective action, we point out that higher order terms in curvature lead to a nonlinear  and non-local relation between the propagating scalar degree of freedom and the Ricci scalar. We show how to resolve these ambiguities and how to obtain the correct action in the Einstein frame. We illustrate our results by looking at $f(R)$ gravity.
\end{abstract}  
\vspace{5cm}

\end{titlepage}



\newpage
\section{Introduction}

Effective field theory techniques represent a powerful tool to deal with quantum gravity. Indeed effective field theory methods enable a decoupling on energy scales. As we are interested in physics below the Planck scale where experiments and observations can be performed, it is sufficient to develop a theory of quantum gravity valid for low energies or equivalently for small space-time curvatures. 
The effective action for quantum gravity can be seen as a series expansion in curvature. Integrating out the graviton fluctuations, one obtains an effective action which to second order in curvature is given by \cite{Weinberg,Donoghue:1994dn,Bar1984,Bar1985,Bar1987,Bar1990,Buchbinder:1992rb,Calmet:2018elv}:
\begin{eqnarray}
S&=& \int d^4x \, \sqrt{-g} \left(  \frac{1}{2}  M_P^2   R + c_1 R^2 + c_2 R_{\mu\nu}R^{\mu\nu}
 - b_1 R \log \frac{\Box}{\mu^2}R -  b_2 R_{\mu\nu}  \log \frac{\Box}{\mu^2}R^{\mu\nu} +{\cal O}(R^3)  \right), \nonumber \\
\end{eqnarray}
where $\mu$ is the renormalization scale. As the theory is not renormalizable in the usual sense, the Wilson coefficients $c_1$ and $c_2$ must be measured in experiments or observations. On the other hand, the Wilson coefficients of the non-local operators $b_1$ and $b_2$ can be calculated from first principles and if the unique effective action formalism is adopted, their values are gauge invariant \cite{Bar1984,Bar1985,Bar1987,Bar1990}. This leads to model independent predictions in quantum gravity which are independent of the ultra-violet completion.

The effective action has a well known potential issue: the term $R_{\mu\nu}R^{\mu\nu}$ leads to a massive spin-2 field which is a ghost, it has an overall minus sign in front of the kinetic term when compared to a standard massive spin-2 field \cite{Stelle:1977ry}.  In this paper, we investigate whether the massive spin-2 ghost is an artifact of the truncation of the action to second order terms in curvature as it is often claimed see e.g. \cite{Alvarez-Gaume:2015rwa}. We point out that the position of the poles due to $R^2$ and $R_{\mu\nu}R^{\mu\nu}$ cannot be affected by operators that are higher order in curvature of the type $R_n.... R_m$ where $R_n$ stands for the Ricci scalar, Ricci tensor or Riemann tensor where we assume that all indices are contracted such that the operator is a scalar. On the other hand, operators of the type $R\Box^n R$ will lead to new poles while shifting the positions of the poles found at second order in curvature. The number of poles grows with the number of operators of the type $R\Box^n R$ that we include in the effective action. Unless these operators can be resummed in a smooth function, the full theory of quantum gravity may contain an infinite number of degrees of freedom.

We will argue that identifying the interactions between the new degrees of freedom requires to map the effective action formulated in the Jordan frame to the Einstein frame. As a byproduct of this work, we show that mapping the effective field theory in the Jordan frame to the Einstein frame can lead to some ambiguities when higher order curvature terms are included in the effective action. While this procedure is well defined for second order curvature terms in the effective action, we point out that higher order terms in curvature lead to a nonlinear  and non-local relation between the propagating scalar degree of freedom and the Ricci scalar. We show how to resolve these ambiguities to obtain the correct action in the Einstein frame and illustrate our results by looking at $f(R)$ gravity.

\section{Degrees of freedom in the gravitational effective action}

In this section, we study the different degrees of freedom in the gravitational effective action. In particular, our aim is to check whether operators of higher order in curvature could affect the poles found at second order in curvature. We will consider a small perturbation given by a symmetric matrix $h_{\mu\nu}$ around a background metric $\bar{g}_{\mu\nu}$. The full metric $g_{\mu\nu}$ is linearized according to
\begin{align}
  g_{\mu\nu} =\bar{g}_{\mu\nu} +\kappa h_{\mu\nu}.\label{metric_definition}
\end{align}
Within this paper we adopt the signature $\operatorname{diag}(+---)$ and define $\kappa$ via the Newton constant as follows:
\begin{align}
  \cfrac{1}{16\pi G} = \cfrac{2}{\kappa^2}.
\end{align}
As $\bar{g}_{\mu\nu}$ defines the geometry of the background spacetime, all indices are raised and lowered with this background metric. 

The definition \eqref{metric_definition} allows one to express all geometric quantities in terms of perturbations $h_{\mu\nu}$. For instance the inverse metric $g^{\mu\nu}$ is given by the following infinite series:
\begin{align}
  g^{\mu\nu} = \bar{g}^{\mu\nu} + \sum\limits_{n=1}^\infty (-\kappa)^n (h^n)^{\mu\nu}=\bar{g}^{\mu\nu} - \kappa h^{\mu\nu} + \kappa^2 h^{\mu\sigma}h_\sigma{}{}^\nu + O(h^2).\label{inverse_metric}
\end{align}
Obviously, the perturbative expansion only makes sense if the perturbation $h_{\mu\nu}$ is small with respect to $\kappa$. 

We denote background quantities by an ``over line''. For instance, as already mentioned $\bar{g}_{\mu\nu} $ is the background metric and $\bar{R}$ is the background Ricci scalar. We denote parts linear in $\kappa$ by a single underline. In such a way $\underline{g}^{\mu\nu} = - \kappa h^{\mu\nu}$ is a part of $g^{\mu\nu}$ linear in $\kappa$. Finally, we denote parts quadratic in $\kappa$ by a double underline, so $\uunderline{g}^{\mu\nu}=\kappa^2 h^{\mu}{}_\sigma h^{\sigma\nu}$ is the part of $g^{\mu\nu}$ quadratic in $\kappa$. These are standard notations, see e.g. \cite{tHooft:1974toh}.

Let us first briefly review the work of Stelle \cite{Stelle:1977ry} for quadratic gravity and identify the ghost due to the term $R_{\mu\nu} R^{\mu\nu}$:
\begin{align}\label{quadratic_action}
  S = \int d^4 x \sqrt{-g} \left[ -\cfrac{2}{\kappa^2} R +c_1 R^2 + c_2 R_{\mu\nu} R^{\mu\nu} \right].
\end{align}
The part of the action quadratic in perturbations describes spin-2 and spin-0 perturbations:
  \begin{align}
    \uunderline{ S}=\int d^4 x \left[ -\cfrac12 ~ h^{\mu\nu} \mathcal{O}_{\mu\nu\alpha\beta} h^{\alpha\beta} \right],
  \end{align}
where the operator $\mathcal{O}$ is given in terms of the spin-2 and spin-0 projectors defined in momentum space:
\begin{align}
  \mathcal{O}_{\mu\nu\alpha\beta} &= -k^2 \left[ \left(1+\cfrac{c_2 \kappa^2 k^2}{2}\right) P^2_{\mu\nu\alpha\beta} - 2 (1-(3c_1+c_2)\kappa^2 k^2) P^0_{\mu\nu\alpha\beta} \right],
\end{align}
where we used the Feynman gauge.
The projection operators $P^{(2)}$ and $P^{(0)}$ are defined as follows:
\begin{align}
  P^{(2)}_{\mu\nu\alpha\beta} &= \cfrac12 \left(\Theta_{\mu\alpha}\Theta_{\nu\beta} + \Theta_{\mu\beta}\Theta_{\nu\alpha}\right) -\cfrac13 ~\Theta_{\mu\nu}\Theta_{\alpha\beta} , \\
  P^{(0)}_{\mu\nu\alpha\beta} &= \cfrac13 ~\Theta_{\mu\nu}\Theta_{\alpha\beta} ,
\end{align}
with $\Theta_{\mu\nu}=\eta_{\mu\nu} - k_\mu k_\nu/ k^2$. Inverting the operator, which requires the introduction of a gauge fixing term \cite{Hindawi:1995an}, yields the propagator in momentum-space with the following gauge invariant part:
\begin{align}\label{localprop}
\bar{D}_{\mu\nu\alpha\beta} = 
\frac{\left(P_{\mu\nu\alpha\beta}^{(2)}-\frac{1}{2}P_{\mu\nu\alpha\beta}^{(0)}\right)}{k^2} 
 -\frac{P_{\mu\nu\alpha\beta}^{(2)} }{k^2 - m_2^2} + \frac{P_{\mu\nu\alpha\beta}^{(0)}}{k^2 - m_0^2} \ ,
\end{align}
where we have used partial fractions to identify the masses of the spin-2 and spin-0 fields
\begin{align}
m_2^2 =\frac{ M_P^2}{-2 c_2 }, \quad m_0^2 =\frac{M_P^2}{4(3 c_1 + c_2 )} \ \ ,
\end{align}
where $M_P=2/\kappa$ is the reduced Planck mass.
The sign in front of the massive spin-2 propagator shows that this field is a ghost \cite{Stelle:1977ry,Hindawi:1995an}, independently on the values of the Wilson coefficients in the effective action.

Our aim is to study whether higher curvature operators such as e.g. $R^3$ or $R_{\mu\nu\alpha\beta} R^{\mu\nu}_{\ \ \rho\sigma} 
R^{\rho\sigma\alpha\beta}$ etc, could affect the position of the poles identified by Stelle \cite{Stelle:1977ry} a long time ago or even remove the ghost from the spectrum of the effective action. To do so, we linearize the effective action for quantum gravity and focus on terms which are second order in the graviton field. We consider a generic term in the effective action of the type
\begin{align}
R_1 R_2 ... R_n=(\overline{R_1}+\underline{R_1}+\uunderline{R_1})(\overline{R_2}+\underline{R_2}+\uunderline{R_2})...(\overline{R_n}+\underline{R_n}+\uunderline{R_n})\ ,
\end{align}
 where $R_j$ stands for the Ricci scalar, Ricci tensor or Riemann tensor and expand it around a background. It is understood that all indices are contracted in an appropriate manner such that these terms are scalars.  Our aim is to show that for $n\ge 3$ there are no terms bilinear in the fluctuations. We first calculate 
\begin{eqnarray}
R_1 R_2&=&(\overline{R_1}+\underline{R_1}+\uunderline{R_1})(\overline{R_2}+\underline{R_2}+\uunderline{R_2})
\\ \nonumber &=& 
\overline{R_1} \ \overline{R_2}+\overline{R_1} \underline{R_2}+\overline{R_1} \uunderline{R_2}+\underline{R_1} \overline{R_2}+\underline{R_1}  \ \underline{R_2}+\uunderline{R_1} \overline{R_2} +{\cal O}(\kappa^3).
\end{eqnarray}
If we evaluate this expression around a flat background, we get $R_1 R_2=\underline{R_1}  \ \underline{R_2}+{\cal O}(\kappa^3)$, these terms are the sources of the poles identified by Stelle. It is clear that if we calculate
$R_1 R_2 ... R_n$, the term $\underline{R_1}  \ \underline{R_2}$ will be multiplied by a $R_j$ evaluated at the background and all bilinear terms will thus vanish. The truncation of the action does not affect the position and nature of the poles identified by Stelle. The same reasoning applies to terms of the type $R_1 \log \Box R_2 ... R_n$, so that the width discussed in \cite{Calmet:2014gya}, will not be impacted by the truncation of the action. 

There is another class of effective operators which, however, can generate new poles and shift the positions of the poles studied by Stelle.
Indeed, terms of the type $R_i \square^n R_j$, when included in the effective action, modify the structure of the propagator. It is straightforward to see this by studying:
\begin{align}
  \uunderline{S}_\text{higher derivatives}=\int d^4 x & \left[ h^{\mu\nu} \left[  a_1 \square + \cdots + a_n \square^n  \right] P^2_{\mu\nu\alpha\beta} ~ h^{\alpha\beta} \right. \nonumber \\
    & \left.+  h^{\mu\nu} \left[  b_1 \square + \cdots + b_n \square^n \right]~P^0_{\mu\nu\alpha\beta}~ h^{\alpha\beta} \right] .
\end{align}
This expression can always be factorized down to the following form:
\begin{align}
  \uunderline{S}_\text{higher derivatives}=\int d^4 x & \left[ h^{\mu\nu} (\square +(m_{0,1})^2) (\square +(m_{0,2})^2) \cdots (\square + (m_{0,n})^2) \square P^2_{\mu\nu\alpha\beta} ~ h^{\alpha\beta} \right. \nonumber \\
    & \left.+  h^{\mu\nu} (\square + (m_{2,1}^2)(\square+(m_{2,2})^2) \cdots(\square +(m_{2,n})^2) \square P^0_{\mu\nu\alpha\beta} h^{\alpha\beta} \right].
\end{align}
In these expression $m_{0,k}$ and $m_{2,k}$ are complex constants determined by the Wilson coefficients $a_k$, $b_k$. Our first observation is that there is an infinite number of poles when the full effective action is considered unless these terms can be resummed in which case, at least in principle, some or even all poles could be eliminated. These poles correspond to massive spin-2 and spin-0 fields. Furthermore, it is easy to see that while the masses of the lightest spin-0 and spin-2 fields discovered by Stelle can be affected, the fact that the effective action contains a ghost cannot be an artifact of the truncation unless again some resummation mechanism is at work. However, this would cast some doubts on the validity of the perturbative expansion\footnote{We note that models with this feature have been considered in e.g. \cite{Biswas:2011ar,Modesto:2011kw,Tomboulis:1997gg,Shapiro:2015uxa}, whether quantum gravity truly leads to such resummable series is an open question.}. Indeed,  all Wilson coefficients of operators of the type $R_i \square^n R_j$ are suppressed by the Planck mass to the $(2n)^{th}$ power. They should not impact the result obtained by Stelle  unless again the perturbative expansion breaks down. We note also that the attempt by Zwiebach in \cite{Zwiebach:1985uq}  of making the ghost non-propagating by adjusting the coefficients of $R^2$ and $R_{\mu\nu}R^{\mu\nu}$ cannot hold as his relation is not invariant under the renormalization group evolution in the case of the effective action \cite{Buchbinder:1992rb}:
 \begin{align}
c_1(\mu)= c_1(\mu_R)- \frac{1}{11520 \pi^2}(3 N_s  - 12 N_f  - 24 N_V-244) \log \left ( \frac{\mu^2}{\mu^2_R} \right)\\
c_2(\mu)= c_2(\mu_R)- \frac{1}{11520 \pi^2}(6 N_s  + 36 N_f  + 72 N_V+1452) \log \left ( \frac{\mu^2}{\mu^2_R} \right).
\end{align}

Another attempt to eliminate the ghost in string theory \cite{Deser:1986xr}, which has the same effective action as the one we considered here, relies on a specific metric transformation due to Kallosh, Tarasov and Tyutin \cite{Kallosh:1978wt} (see also \cite{Deser:1986xr}). Indeed transformations of the type $g^\prime_{\mu\nu} \to g_{\mu\nu} + aR_{\mu\nu}+ b g_{\mu\nu} R$ can be used to eliminate the terms $R^2$ and $R_{\mu\nu} R^{\mu\nu}$ in the action. However, it has recently been argued \cite{Modesto:2017hzl,Goncalves:2017jxq}  that such transformations are not appropriate to eliminate the ghost degree of freedom. In our case, as we need to rely on the unique effective action \cite{Bar1985} to insure that the Wilson coefficients, and thus observables, are gauge invariant, the coefficients of $R_{\mu\nu}R^{\mu\nu}$ and $R^2$ are uniquely fixed and non-zero.

The ghost associated  with $R_{\mu\nu}R^{\mu\nu}$  has been much discussed in the literature and while it is often argued that it is pathological because it carries negative energy, we are not aware of a specific calculation showing what could go wrong in an actual physical setting. For example, the emission on the massive spin-2 waves corresponding to this ghost by a binary system such as two black holes does not lead to any pathology \cite{Calmet:2018qwg,Calmet:2018rkj}. Quantum gravitational corrections to metric describing black holes or stars are well defined \cite{Calmet:2018elv,Calmet:2017qqa}.  We note however that there could be a simple explanation for this observation. The action of effective field theory for quantum gravity is a classical field theory as the quantum fluctuations of the graviton have been integrated out. The ghost field does not need to be quantized, there is thus no issue with unitarity. Secondly, the ghost is unstable as it will decay quickly to gravitons and matter fields \cite{Calmet:2014gya}.  Indeed, it can be shown that the non-local terms lead to a width for the massive spin-2 ghost. Furthermore, when the local part of the Einstein frame effective action  is linearized one obtains
\begin{eqnarray}
S=\int d^4 x \left[\left (- \frac{1}{2} h_{\mu\nu} \Box h^{\mu\nu}
 +\frac{1}{2} h_{\mu}^{\ \mu} \Box h_{\nu}^{\ \nu}  -h^{\mu\nu} \partial_\mu \partial_\nu h_{\alpha}^{\ \alpha}+ h^{\mu\nu} \partial_\rho \partial_\nu h^{\rho}_{\ \mu}\right) \right. \\ 
\left. \nonumber -\left ( -\frac{1}{2} k_{\mu\nu} \Box k^{\mu\nu}
 +\frac{1}{2} k_{\mu}^{\ \mu} \Box k_{\nu}^{\ \nu}  -k^{\mu\nu} \partial_\mu \partial_\nu k_{\alpha}^{\ \alpha}+ k^{\mu\nu} \partial_\rho \partial_\nu k^{\rho}_{\ \mu}
 \right.  \right.  \\ \left. \left. \nonumber
 -\frac{M_2^2}{2} \left (k_{\mu\nu}k^{\mu\nu} - k_{\alpha}^{\ \alpha} k_{\beta}^{\ \beta} \right )
 \right)  \right.
  \\
  \nonumber
 \left. + \frac{1}{2} \partial_\mu \sigma  \partial^\mu \sigma
  - \frac{M_0^2}{2} \sigma^2 - \sqrt{8 \pi G_N} (h_{\mu\nu}+k_{\mu\nu}+\frac{1}{\sqrt{3}} \sigma \eta_{\mu\nu})T^{\mu\nu} 
  \right ],
\end{eqnarray}
which leads to the same field equations as
\begin{align}
S=\int d^4 x \left[\left (- \frac{1}{2} h_{\mu\nu} \Box h^{\mu\nu}
 +\frac{1}{2} h_{\mu}^{\ \mu} \Box h_{\nu}^{\ \nu}  -h^{\mu\nu} \partial_\mu \partial_\nu h_{\alpha}^{\ \alpha}+ h^{\mu\nu} \partial_\rho \partial_\nu h^{\rho}_{\ \mu}\right) \right. \\ 
\left. \nonumber +\left ( -\frac{1}{2} k_{\mu\nu} \Box k^{\mu\nu}
 +\frac{1}{2} k_{\mu}^{\ \mu} \Box k_{\nu}^{\ \nu}  -k^{\mu\nu} \partial_\mu \partial_\nu k_{\alpha}^{\ \alpha}+ k^{\mu\nu} \partial_\rho \partial_\nu k^{\rho}_{\ \mu}
 \right.  \right.  \\ \left. \left. \nonumber
 -\frac{M_2^2}{2} \left (k_{\mu\nu}k^{\mu\nu} - k_{\alpha}^{\ \alpha} k_{\beta}^{\ \beta} \right )
 \right)  \right.
  \\
  \nonumber
 \left. + \frac{1}{2} \partial_\mu \sigma  \partial^\mu \sigma
  - \frac{M_0^2}{2} \sigma^2 - \sqrt{8 \pi G_N} (h_{\mu\nu}-k_{\mu\nu}+\frac{1}{\sqrt{3}} \sigma \eta_{\mu\nu})T^{\mu\nu} 
  \right ],
\end{align}
which must thus describe exactly the same physics (as the field equations are the same, assuming identical boundary conditions), but here the massive spin-2 $k_{\mu\nu}$ is not a ghost, it simply couples to $T^{\mu\nu}$ with the negative Planck mass.  It simply leads to a repulsive force. Any pathology, if any, must be due to non-linear effects, however these effects are essentially irrelevant for any practical calculations in cosmology or astrophysics as all data we have is at best sensitive to linear gravitational effects.

Either the ghost in the effective action is harmless (which is compatible with the Lee-Wick interpretation pushed in \cite{Modesto:2017hzl,Goncalves:2017jxq}) or it is problematic for any theory of quantum gravity including string theory. Indeed it was shown in \cite{Alvarez-Gaume:2015rwa}, that whatever compactification is chosen in string theory, one ends up with such terms in the effective action.  It is often argue that the ghost is an artifact of the truncation of the effective action and higher curvature terms could remove the ghost, see e.g. \cite{Alvarez-Gaume:2015rwa}. We have shown that this is not the case unless a very specific resummation takes place. We have identified a class of higher dimensional operators that introduce new poles in the action but the fact position of the pole corresponding to the massive spin-2 ghost cannot be altered by higher dimensional operators unless these higher curvature operators have extremely large Wilson coefficients. In any case, although the position of poles might be shifted, the existence of these poles cannot be altered. Finally, we note that higher derivative scalar field theories have been considered by Hawking and Hertog\cite{Hawking:2001yt}, who reached the conclusion that theories with ghosts can make sense.   In the next section, we study the interactions among the new gravitational degrees of freedom contained in quantum gravity. The interactions are needed to probe quantum gravity in its nonlinear regime. 

\section{Interactions}
While it is straightforward to obtain the masses of the fields present in the effective action by looking at the poles of the metric propagator in the Jordan frame, identifying their interactions requires to map the theory to the Einstein frame. This is trivial in the case of the quadratic theory considered by Stelle. However, things are more complicated when higher curvature terms are considered. We will illustrate this complication using the well studied $f(R)$ gravity model. To do so we revisit the usual transformations involved in the map from the Jordan frame to the Einstein frame.

One starts from an $f(R)$ action in the Jordan frame.
\begin{align}
  S&= -\cfrac{2}{\kappa^2} \int d^4 x \sqrt{-g} f(R). \label{action_A} 
\end{align}
It is well known that this action describes two fields, a massless spin-2 field $h_{\mu\nu}$ and massive scalar field $\phi$. To identify these fields, we could expand the metric $g_{\mu\nu}$ around a background according to $g_{\mu\nu}=\bar g_{\mu\nu}+h_{\mu\nu}+\phi\eta_{\mu\nu}$. While one can easily obtain the masses of the fields by looking at the positions of the poles of the Green's function, identifying the scalar potential requires mapping the action to the Einstein frame.  To do so, one performs a field redefinition given by
\begin{align}
g_{\mu\nu} = \Omega^2 \tilde{g}_{\mu\nu}, 
\end{align}
where
\begin{align}
  \ln\Omega=\frac{1}{2} \ln\Omega^2 =  -\frac{1}{2}\ln f'(R).
\end{align}

This field redefinition leads to the following relations:
\begin{align}\label{conformal_mapping}
  \begin{cases}
     \sqrt{-g} = \Omega^4 \sqrt{-\tilde{g}}, \\
    R = \Omega^{-2} \left[\tilde{R} -6 \square \ln \Omega -6 (\nabla\ln\Omega)^2 \right] .
  \end{cases}
\end{align}
Finally, the action can be expressed in terms of the new metric $\tilde{g}$ and a scalar $\phi$
\begin{align}\label{action_D}
  S= \int d^4 \sqrt{-\tilde{g}} \left[-\cfrac{2}{\kappa^2} \tilde{R}  +\cfrac12 (\nabla \phi)^2 - V(\phi)\right].
\end{align}
Here $\phi$ is defined in terms of $R$ as follows:
\begin{align}\label{phi_definition}
  \phi=\cfrac{\sqrt{6}}{\kappa} \ln f'(R).
\end{align}
The potential $V(\phi)$ is given by the following expression:
\begin{align}
  V(\phi) = \cfrac{2}{\kappa^2}~ \cfrac{f(R(\phi)) - \sigma(\phi) f'(R(\phi))}{[f'(R(\phi))]^2}.
\end{align}
We immediately see that the relation between $R$ and $\phi$ is not necessarily unique. For example, if we consider $f(R)=R^n$ with $n\ge 3$, then equation (\ref{phi_definition}) has multiple roots. The relation between $\phi$ and $R$ is not uniquely defined and thus neither is the potential. 

The simplest $f(R)$-gravity model containing a scalar field which demonstrates the issue is given by
\begin{align}\label{action_2}
  S = -\cfrac{2}{\kappa^2} \int d^4 x \sqrt{-g} \left[ R- \cfrac{1}{6m^2} R^2 +\cfrac{1}{\varepsilon^4} R^3 \right].
\end{align}
The relation between field $\phi$ and $R$ is not defined uniquely. The two solutions for $R(\phi)$ are given by:
\begin{align}
  R(\phi)_\pm= \cfrac{\varepsilon^4}{18 m^2} \pm\cfrac{\varepsilon^4}{6} \sqrt{\cfrac{1}{9m^4} + \cfrac{12}{\varepsilon^4}\left(\exp\left[\cfrac{\kappa\phi}{\sqrt{6}}\right]-1\right) }.
\end{align}
while the two solutions are a priori acceptable, only one corresponds to the field configuration present in the original $f(R)$-gravity action. To identify the correct solution, we look at the potential in the Einstein frame action and calculate the masses of the two solutions according to
\begin{align}
  m^2_\text{scalar} = \left. \cfrac{d^2}{d\phi^2} V(\phi) \right|_{\phi=0}.
\end{align}
We find
\begin{align}
  \begin{cases}
    \left. m^2_\text{scalar} \right|_{R_-} = m^2 ,\\
    \left. m^2_\text{scalar} \right|_{R_+} = -m^2 + \cfrac{\varepsilon^4}{9 m^2} - \cfrac{2 \varepsilon^8}{2187 m^6} .
  \end{cases}
\end{align}
The formal expression \eqref{action_2} admits $\varepsilon\to\infty$ limit, therefore all physical quantities such as the scalar field mass must admit the same limit. The scalar field mass obtained using the solution $R_+$ diverges in $\varepsilon \to \infty$ limit, so this solution is unphysical. We see that when mapping a gravitational model from the Jordan frame to the Einstein frame, it is crucial to select the solution for the scalar field that reproduces the correct position of the pole. The same observation applies when considering models involving the Riemann tensor. In that case the spectrum of the model involves massive spin-2 fields \cite{Stelle:1977ry,Hindawi:1995an}.

\section{Conclusions}

In this paper we have considered the effective field theory for quantum gravity and studied its particle spectrum. We have shown that quantum gravity contains potentially an infinite number of degrees of freedom each corresponding to a pole in the metric Green's function unless a very specific resummation takes place. We have shown that special care needs to be taken when identifying the interactions between these new degrees of freedom. This requires mapping the theory from the Jordan frame to the Einstein frame and we have discuss how to perform this map correctly when higher order curvature terms are present. We have shown that the truncation of the action is not the reason for the presence of a ghost in the spectrum of the theory. Either this ghost is harmless or all theories of quantum gravity are in trouble as there is no obvious way to remove this ghost from the spectrum of quantum gravity.

{\it Acknowledgments:}
The work of XC is supported in part  by the Science and Technology Facilities Council (grant number ST/P000819/1). 


\bigskip{}

\end{document}